\documentclass[british,a4paper]{article}
\usepackage[T1]{fontenc}
\usepackage[utf8]{inputenc}
\usepackage{lmodern}

\usepackage{babel}
\usepackage{hyperref}
\usepackage{microtype}
\usepackage{authblk}
\usepackage{graphicx}
\usepackage{subcaption}
\usepackage{epstopdf}
\usepackage{amsmath, amssymb, amsfonts, amsthm, mathrsfs}
\usepackage{float}
\usepackage{booktabs}
\usepackage{siunitx}
\usepackage{raccourcis}

\begin{document}

\title{The $N$-achromat and beyond:
a unified variational framework for optimal chromatic aberration correction.}
\author[1]{Bastien Laville}
\author[1]{Benjamin Aymard}
\affil[1]{Thales Alenia Space, 5 Allée des Gabians, 06150 Cannes, France}
\maketitle

\begin{abstract}
  In this article, we present novel and effective methods for reducing chromatic aberrations in
  cemented lens systems. We derive an analytical solution coined the pentachromat, which
  corrects five distinct colors. This method can naturally be extended to accommodate
  an arbitrary number of lenses and to correct for a customized selection of spectral lines.
  Since correcting for specific rays rather than the entire residual spectrum
  can overconstrain the system, we introduce a variational formulation. This approach
  tames the residual spectrum by several orders of magnitude compared to conventional
  designs like the superachromat, while giving theoretical guarantees to reach the
  optimal solutions.
  Furthermore, this innovative methodology opens up previously uncharted design
  possibilities, such as multiple-focal-length achromatic systems. This allows for the
  selection of specific optical powers paired with desired bandwidths, enabling the
  design of highly specialized and tailored optical systems.
  Finally, we couple our variational framework with a combinatorial search, allowing to
  find the type of glasses and their geometry such that it reaches the best residual spectrum
  over an available catalogue.
\end{abstract}
\section{Introduction}

\emph{How many lenses does one need to make a camera?}
The first records of a primitive imaging systems date back to Antiquity,
a period that witnessed the creation, by different inventors, and
approximately at the same time,
of the pinhole camera (from Latin, the "dark room").
The device consists of a closed box with a hole allowing the light to
enter on one side and form an image on the other.
Despite its simplicity, it has proven to be very useful in
imaging the Sun.
This is the simplest imaging system, and it contains no lens,
even though the optical propeties of lenses were already known in Antiquity.
\footnote{
  one may think of the famous Nimrud lens
  (710 BC, Assyria, that was possibly used for optical purpose not
  confirmed as of today),
  the burning stone cited in Aristophane's "the clouds" (423 BC),
  or magnifying filled jar mentionned in Seneca's "Quaestiones
  Naturales" (50 AD).
}

Lenses were properly introduced in imaging systems only centuries later,
when Zacharias Janssen used a polished piece of glass in order to
focus light to observe tiny objects:
it was the birth of microscopy (1590).
Soon after, the system was adapted for distant objects imaging,
notably by Kepler, and then Galileo,
leading to the invention of the refracting telescope
Together with this novelty, came the first problem in lens design:
the phenomenon of chromatic aberrations.
Different colors will focus at different distances of the lens
system, degrading the image,
with rainbow fringes appearing on the edges of white objects.
Famous scientists at the time thought that chromatic aberration was
impossible to
correct, including Newton\footnote{
  \emph{"But by reason of this different refrangibility, I do not yet
    see any other mean
    of improving telescopes by refractions alone than that of
  increasing their lengths "}
  (Opticks, 1704).
  as he thoughts that all glasses have the same refraction index.
  This assertion was proven to be erroneous, notably when Frauhoffer
  introduced a method to measure this index,
  based on the lines he had previously discovered.
} himself.

The proper solution\footnote{A famous solution to reduce chromatic
  aberrations was the aerial
  telescopes (Hevelius, Huygens), used notably by Cassini at Paris
  observatory to
  discover Tethys in 1684. However, it was not a real correction, but
  rather a reduction, of the
chromatic aberration.} came decades after, when a London barrister,
Chester Moore Hall, proposed the achromatic doublet (1729):
combining the classical crown glass (used for instance in cathedral
stained glass windows during middle age),
with the more recently discovered flint glass (Ravenscroft, 1674).
The doublet was able to focus two different colors at the same focal point.
Decades after, John Dollond used this solution to make the first
achromatic refracting telescope (1758).
In 1879, Ernst Abbe, working at the time in the field of microscopy,
managed to produce the first apochromatic microscope:
with three well-chosen glasses, he showed that it was possible to
correct exactly three colors, as a direct extension of the method
used for the doublet.
One key part in his solution was the use of what was called "glass X"
at the time, a glass made of fluorite, with a very specific chromatic
dispersion.
The neverending quest for precision led Hertzberger to design in 1963
the superachromat lens.
Using four lenses, this design was able to go further than the
apochromatic design,
by giving a theoretical formula correcting four colors simultaneously.

As we have seen along this chronology,
correcting chromatic aberrations has always been a major concern in
lens design. The
reduction of chromatic error is a critical necessity across several
scientific and
technological domains: in photography and cinematography, robust
correction is essential for
achieving the
high-fidelity color and sharp detail required by modern sensors.
Biomedical imaging and
diagnostics rely on rigorous chromatic error correction to prevent
color shifts from
creating artifacts and to facilitate accurate pathological assessment
in tools like
fluorescence microscopy \cite{Betzig2006}, sometimes algorithmtically
catered for. We could also cite lithography and
semiconductor manufacturing,
even with a spatially coherent laser\footnote{even 'very narrow'
  bandwidths (sub‑picometer) cannot always be assumed negligible when
  numerical apeture increases (as with the current trend) and lens
materials are constrained} there is room for degradation
due to chromatic aberration \cite{Kroyan2000}. One could finally quote remote
sensing and
satellite imaging to provide high-quality, spectrally accurate data.

In this article, we propose original designs bringing further the
correction of chromatic
aberration.
First, we clarify the historical approach,
and present a simple way to solve the problem of correcting exactly $N$ colors.
However, it turns out that there might be more efficient approaches
in lens design.
As the number of lenses $N$ increases, the residual spectrum might
behave widly between
the $N$ corrected colors.
We propose an original point of view: rather than correcting exactly
$N$ prescribed wavelengths,
we suggest to minimize directly the residual chromatism deviation on
an entire spectral
window. This avoids to overconstrain the problem and is more
efficient, as the roots (aka
wavelengths corrected) are computed in a way that minimises the chromatic error.
We define this system through a method of constraint optimisation
\cite{Lagrange2009,Karush2013,Kuhn1951}
framework called the
Kash-Kuhn-Tucker (KKT) system. Then, we explore how to compute in a
stable fashion the curvature
values $K_i$ yielding the miniminal chromatic deviations from the
target power, formulating
the problem with a Chebyschev expansion or by considering a recast such
as the null-space method \cite{Gould2001}. Our method is able to
compute the best
$K_i$ with minimimal
chromatic abberation on a windows of wavelength, which paves the way
to system with less
glasses than $N$, while keeping a
competitive performance with capable $N$-achromat. In addition to the
latter, our method
is not limited to fix target power $\phi_0$ but it also caters for
even varying with
$\lambda$ target power, or with separated windows for example in
near-UV and near-IR.

Finally, not only do we optimise on the curvature of the
glasses $K_i$ but also on the type of glasses themselves: leveraging the Schott
catalogue, we give a full-fledged method which finds a $N$-optical
system with the smallest
chromatic error, given only the number of lenses $N$ and a catalogue
of available
lenses. This overall work provides a powerful new tool for designers that
not only finds mathematically optimal solutions but also helps
navigate practical
constraints like manufacturability.

\section{From superachromat to $N$-achromat}

In this section, we present a general method to design an optical
system that corrects $N$ colors simultaneously,
starting with a reminder abouth chromatic aberrations.
The interested reader is refered to \cite{BornWolf1999} and
\cite{Smith2007} for a gentle
introduction to these concepts.
Concerning more thorough and advanced explanations, including the
design of apochromat
and superachromat, the
reader is refered to \cite{Sasian2017}.

\subsection{Notions on chromatic aberration}

Let us recall the surface imaging formula for one refractive surface.
Given a light wavelength $\lambda \in \R$, one gets:

\begin{equation}
  \frac{n_1(\lambda)}{s} + \frac{n_2(\lambda)}{s'} =
  \frac{n_2(\lambda) - n_1(\lambda)}{R},
  \label{Descartes}
\end{equation}

relating an object position $s$, its image position $s'$ by the
surface of curvature radius $R > 0$,
and refractive indexes $n_1, n_2 > 0$ of the two media.
The latter is often computed using the Sellmeier formula \cite{Sellmeier1871a}:

\begin{equation}
  n(\lambda)^2 =  1 + \frac{B_1 \lambda^2}{\lambda^2 - C_1} +
  \frac{B_2 \lambda^2}{\lambda^2 - C_2} + \frac{B_3 \lambda^2}{\lambda^2 - C_3},
  \label{eq:sellmeier}
\end{equation}

with experimental coefficients $B$ and $C$, specific to the
considered glass. Loosely
speaking, it holds from 0.2$\mu$m to 2$\mu$m.
A thin lens is made of two such spherical surfaces, surrounded by
air. Using the latter
formula, one yields:

\begin{align*}
  \frac{1}{s} + \frac{n(\lambda)}{s''} &= \frac{n(\lambda) - 1}{R_1},\\
  \frac{n(\lambda)}{d-s''} + \frac{1}{s'} &= \frac{1 - n(\lambda)}{R_2}.
\end{align*}

Assuming that the length is thin $d \approx 0$, and summing both
expressions, one reaches:

\[
  \frac{1}{s} + \frac{1}{s'}
  =
  (n(\lambda)-1)\left(\frac{1}{R_1} -  \frac{1}{R_2}\right).
\]

Considering an object at infinity $s = +\infty$, one obtains the
lensmaker equation,
giving the power (inverse of the focal length F) of the lens:

\[
  \phi(\lambda) = \frac{1}{F} =
  \underbrace{(n(\lambda)-1)}_{\mbox{ Physics }}
  \underbrace{\left(\frac{1}{R_1} -  \frac{1}{R_2}\right)}_{ \mbox{
  Geometry } }.
\]

Classically, we denote by $K$ the geometric factor, defined as:

\begin{align}
  K = \left(\frac{1}{R_1} -  \frac{1}{R_2}\right).
  \label{eq:K}
\end{align}

In the case of one thin lens, we remark that two different wavelengths
will give two different optical powers leading to two different focal lengths.
This effect is called the longitudinal chromatic aberration (LCA):
\[
  \mbox{LCA} = |\phi(\lambda_2) - \phi(\lambda_1)|.
\]

Given an object of size $y$, the size $y'$ of its image is given by:

\[
  y' = m y,
\]

with $m$ the magnification defined as:

\[
  m = \frac{s'}{s} = \frac{1}{\phi s - 1}.
\]

As this quantity is wavelength dependant, then the object size also is.
Transverse chromatic aberration (TCA) is precisely the variation of object
size with respect to wavelength:

\[
  \mbox{TCA} = |y'(\lambda_2) - y'(\lambda_1)| = |m(\lambda_2) -
  m(\lambda_1)| \times |y|.
\]

If the LCA is perfectly corrected, then so is the TCA.
Finally, let us recall the definition of the residual chromatic
aberration (RCA),
the deviation of focus on a spectral window $[\lambda_1,\lambda_2]$
from a target optical
power $\phi_0 \in \R^*$, as:

\begin{equation}
  \mbox{RCA} = \int_{\lambda_1}^{\lambda_2} \left(\phi_0 -
  \phi(\lambda) \right)^2 \ud \lambda.
  \label{RCA}
\end{equation}

Reducing the RCA is more general than reducing LCA, and therefore,
a minimal RCA yields low LCA and low TCA. In the following, we will use the
RCA as our base metric to benchmark the various achromats.

\subsection{A $N$-achromat design \label{sec:n-achromat}}

Using surface imaging formula iteratively,
one can show that the total power of a cemented system,
for which $N \in \N^*$  thin lenses are put in contact, is:

\begin{equation}
  \phi(\lambda)
  = \sum_{j=1}^N \phi_j(\lambda)
  = \sum_{j=1}^N (n_j(\lambda)-1) K_j.
  \label{power}
\end{equation}

A basic $N$-achromatic lens corrects LCA on several spectral lines:
it brings $N$ distincts colors to the same focal point $F =
\frac{1}{\phi_0}$, and is therefore solution to the linear system:

\begin{equation}
  \begin{pmatrix}
    (n_1(\lambda_1) - 1) & \dots & (n_N(\lambda_1) - 1) \\
    \dots & \dots & \dots \\
    (n_1(\lambda_N) - 1) & \dots & (n_N(\lambda_N) - 1)
  \end{pmatrix}
  \begin{pmatrix}
    K_1 \\
    \dots \\
    K_N
  \end{pmatrix}
  =
  \phi_0
  \begin{pmatrix}
    1 \\
    \dots \\
    1
  \end{pmatrix}.
  \label{N_achromat}
\end{equation}

\begin{remark}
  Historically, the case $N=2$ has been called an achromatic doublet
  (Chester Moore-Hall 1729, John Dollond, 1758),
  the case $N=3$ an apochromatic lens (Abbe, 1879),
  the case $N=4$ a superachromatic lens (Hertzberger, 1963,
  \cite{Hertzberger1963}).
\end{remark}

Let us remark that the matrix encodes light-matter interaction
properties of the system,
via a choice of glass and wavelengths,
whereas the unknown vector represents its geometry, via the lenses curvatures.
As long as the wavelengths, but also the glasses, are pair-wise distincts,
then the matrix is invertible, and herefore the system admits a unique solution.

\begin{remark}
  In this work, we do not rely on the well-known Abbe number $V$.
  The latter appears naturally when one wants to design an achromat doublet:
  \begin{align*}
    \phi(\lambda_F)
    &\approx \phi(\lambda_C) + (\lambda_F - \lambda_C) \, \frac{\partial
    \phi}{\partial \lambda}\\
    &= \phi(\lambda_C) + (\lambda_F - \lambda_C)K \, \frac{\partial
    n}{\partial \lambda} \\
    &= \phi(\lambda_C) + \frac{\phi(\lambda_d)}{V},
  \end{align*}

  with Abbe number

  \[
    V = \frac{n_d - 1}{n_F - n_C}.
  \]

  This concept is relevant especially when one considers the
  Fraunhoffer $C$, $d$ and $F$ lines.
  The definition can be generalized to three arbitrary chosen spectal lines,
  still, it is based on an approximation of $\frac{\partial n
  }{\partial \lambda}$.
  Hence, we chose to go back to the original formulation as it is more general.
  In the examples, we will
  consider the visible spectrum (Vis) + Near Infrared (NIR) and will
  rely on the Sellmeier formula, but
  note that this is easily generalisable to any bandwith where
  $n(\lambda)$ is known.
\end{remark}

Solving the linear system using symbolic computation gives access to
explicit, exact solutions.
To do so, one possibility is to use the LU factorisation algorithm
\cite{Banachiewicz1938}, and use this decomposition to solve exactly the system.
Once the geometric factors $K_i$ are known, the optical designer can
choose the radiuses of
curvatures for each lens that best fit his problem using \ref{eq:K}.

\begin{remark}
  An alternative approach when one wants to solve the problem of achromatism,
  is to carefully chose the glasses in the system. Several techniques
  exists, such as PV diagrams.
  We refer the reader to \cite{CarneirodeAlbuquerque:12} for a recent
  work in the field.
  In this article our approach is the following: first, we provide a
  solution, given a set of glasses,
  an then, we either resort on a brute force approach, looping over all the
  available glasses, and a faster beam search method.
\end{remark}

An \emph{achromat} is generally understood as a system fulfilling the
condition of
achromatism, namely zeroing out the LCA for $N$ chosen colors.
However, the interesting quantity here is there RCA,
and there is room for improvement as the achromatism condition can be
overconstraining.
In the following, we relax this condition and form an
optimal \emph{achromatic} design, in the sense that it minimises the
RCA on a bandwith rather than
perfectly annulating the LCA on $N$ spectral lines.
\section{The optimal achromatic design for $N$ lenses}

\subsection{A variational insight}

Historically, achromatic lenses were designed to zero out the power
error at specific, discrete wavelengths. This approach, centered on
finding the roots of a polynomial representation of the chromatic
error, was logical when optical designers focused on the performance
for a few precise spectral lines.

However, constraining the polynomial's roots can be overly restrictive.
While it guarantees perfect correction at selected wavelengths, it
leaves no degrees of freedom to minimise other important performance
metrics, such as the
RMS (root mean-square) or peak-to-valley power errors across a
continuous spectral band $(\lambda_1, \lambda_2)$. In layman
terms, specifying the roots to ensure perfect achromatism at certain
wavelengths can overconstrain the problem and results in a $\phi(\lambda)$ which
varies widly on a bandwith of interest $(\lambda_1, \lambda_2)$.
This is related to the dichotomy encountered in polynomial fitting,
between \emph{interpolation} (roots fiding) and \emph{approximation}
(a metric minimisation).

In this work, we shift the paradigm. Instead of forcing the error to
be zero at discrete points, we propose to place the minimisation of a
metric at the core
of the design. This approach seeks to fulfill a set of physical
constraints while
simultaneously minimizing the residual spectrum\footnote{integrated
  quadratic difference between
the system's power and its target on a specific bandwidth.} and a
target power, $\phi_0$,
over the entire bandwidth of interest. Hence, it is \emph{optimal} in
the sense that it
computes the geometric factor $K$ yielding the smallest chromatic deviations.

\subsection{The Lagrange formulation \label{sec:kkt}}

We now formulate the design problem within the framework of
constrained optimisation. This allows us to balance the primary goal
of minimizing chromatic aberration with mandatory physical constraints.

Consider a system of $N \in \N^*$ lenses. The primary constraint is
the focal constraint, namely that the sum of the individual lens
powers, $\phi_i$, must equal a target power $\phi_0 \in \R^*$:

\begin{align}
  \sum_{i=1}^N \phi_i &= \phi_0.
  \label{eq:sum-power}
\end{align}

Our objective is to make the system's power function,
$\phi(\lambda)$, as close as possible to a target function,
$t(\lambda)$, across the spectral band $(\lambda_1, \lambda_2)$. For
simplicity in this formulation, we consider a constant target power,
$t(\lambda) = \phi_0$. More precisely, we want to minimise the RCA namely
residual chromatic aberration/error/deviations presented above in
Equation \eqref{RCA} :

\begin{align}
  \mathrm{RCA}_{(\lambda_1, \lambda_2)} =
  \int_{\lambda_1}^{\lambda_2}
  \left(\phi(\lambda) - t(\lambda) \right)^2 \, \mathrm{d} \lambda.
  \label{eq:mse}
\end{align}

For the sake of generalisability, we may also refer to it as the MSE
(for mean-square error)
as it is a term widely used in physics, but in this context RCA and
MSE can used interchangeably.
Now, recall that the optical power at a wavelength $\lambda$ can be written as

\[
  \phi(\lambda) = \sum_{i=1}^N K_i  \left(n_i(\lambda) - 1 \right),
\]

where $n_i(\lambda)$ is the refractive index of lens $i$ and $K_i$ is
a geometric factor for lens $i$ (related to its
curvatures). The problem is now clear, as it boils down to:
\emph{"find the set of coefficients $\{K_i\}_{i=1}^{N}$ that
  minimises the residual chromatic error in Equation \ref{eq:mse}
  while satisfying the
power summation constraint"}.

This is a classic problem in calculus of variations that can be solved
using the method of Lagrange multipliers \cite{Lagrange2009}. The
method transforms a
constrained optimisation problem into an unconstrained one by
introducing an additional variable, the Lagrange multiplier, for each
constraint. Considering problems with only equality constraints, this is a
direct application of the Lagrange multiplier theorem; when
inequality constraints are also present, the framework extends to the
more general Karush-Kuhn-Tucker (KKT) conditions \cite{Karush2013,
Kuhn1951}. In our case, with a
single equality constraint in Equation \eqref{eq:sum-power}, the KKT
conditions simplify to the Lagrange conditions  \cite{Lagrange2009}.



To solve this numerically, we first discretize the problem. We
evaluate the power at $P$ sampled wavelengths,
$\{\lambda_k\}_{k=1}^P$, within the band of interest. The integral in
the RCA is thus replaced by a sum.
We can decompose the power at each sampled wavelength,
$\phi(\lambda_k)$, into its mean component and its higher-order deviations:

\[
  \phi(\lambda_k) = \underbrace{\bar\phi}_{\text{mean}} +
  \underbrace{\phi_\mathrm{high}(\lambda_k)}_{\text{high-order}}.
\]

The constraint in Equation \ref{eq:sum-power} fixes the mean power, $\bar\phi
= \phi_0$. Therefore, minimizing the RCA is equivalent to minimizing the
variance of the power function, which is the sum of the squared
high-order deviations. The optimisation problem becomes:

\[
  \min_{K} \sum_{k=1}^P \phi_\mathrm{high}(\lambda_k)^2
  \quad \text{subject to} \quad \bar\phi = \phi_0.
\]

Let us express this in matrix form. Let $D_\mathrm{high} \in \R^{N \times
P}$ be the matrix where each entry $D_\mathrm{high}^{(i,k)}$ contains
the high-order contribution of lens $i$ at wavelength $\lambda_k$. The
objective function can then be written as a quadratic form:

\[
  \sum_{k=1}^{P} \phi_\mathrm{high}(\lambda_k)^2
  = \sum_{k=1}^{P} \left( \sum_{i=1}^{N} K_i \,
  D_\mathrm{high}^{(i,k)} \right)^2
  = K^\top \underbrace{D_\mathrm{high} D_\mathrm{high}^\top}_{S} K,
\]

where $K$ is the vector of coefficients $K_i \in \R^N$, and $S = D_\mathrm{high}
D_\mathrm{high}^\top$ is an $N \times N$ symmetric matrix. Each element
$S_{ij}$ of this matrix represents the correlation between the
high-order chromatic contributions of lens $i$ and lens $j$ across the
sampled spectrum.



Minimizing the quadratic form $K^\top S K$ subject to a linear equality
constraint is a standard problem that leads to the KKT
system\footnote{out of generalisation we call it the KKT system,
  but note that we rely only on equality constraint and therefore
should we call it Lagrange system.} of
equations. The core idea is to build a \emph{Lagrangian} function
$\lagr(K, \lambda)$ that incorporates both the objective function
and the constraint:

\begin{align}
  \lagr (K, \lambda) = K^\top S K - \lambda \left(C_0^\top K -
  \phi_0 \right),
  \label{eq:lagrange}
\end{align}

where $C_0 = (1, \dots, 1)^\top$ and the constraint in Equation
\eqref{eq:sum-power} is written as $C_0^\top K = \phi_0$.
Now, the only equality constraint \eqref{eq:sum-power} is a linear
one, so stationarity, feasability and constraint qualification
\cite{Kuhn1951} are satisfied. Therefore, a Lagrange multiplier
always exists, and if the matrix $S$ is positive definite (meaning not
eigenvalue is equal to zero), then $K^\top S K$ is strongly convex.
Hence, a minimiser $K^*$ exists and it is unique.
The optimal solution $K^*$ is found at a stationary point of the
Lagrangian, where the gradient with respect to both $K$ and the
multiplier $\lambda$ is
zero. This yields a system of linear equations:

\begin{align}
  \begin{pmatrix}
    2 S + \varepsilon I & C_0 \\
    C_0^\top & 0
  \end{pmatrix}
  \begin{pmatrix}
    K \\ \lambda
  \end{pmatrix}
  =
  \begin{pmatrix}
    0 \\ \phi_0
  \end{pmatrix}.
  \label{eq:KKT}
\end{align}

To be more specific:
\begin{itemize}
  \item $2S$ is the Hessian of the quadratic objective function;
  \item $C_0$ and $C_0^\top$ represent the linear equality constraint from
    Eq. \ref{eq:sum-power};
  \item $\lambda$ is the Lagrange multiplier associated with this constraint;
  \item $\varepsilon > 0$ is a small Tikhonov regularization term,
    ($\varepsilon I$) added to the Hessian to ensure the matrix is
    well-conditioned and to improve the numerical stability of the solution.
\end{itemize}

Solving this block matrix system yields the optimal lens contributions
$K^*$ that minimise the chromatic deviations while strictly satisfying the
total power constraint. The KKT framework thus provides a powerful
generalization of the classical normal equations used in unconstrained
least-squares problems. While an unconstrained minimisation would simply
solve $S K^* = 0$, the KKT system augments this to enforce physically
meaningful constraints.

\begin{remark}
  The KKT framework carry out not only the equality constraint such
  as Equation \eqref{eq:sum-power}, but also \emph{inequality
  constraints}, for it could modelise physical constraint such as
  $K_\text{min} < K < K_\text{max}$. To enforce this type of
  inequality, one would have to add another Lagrange multiplier $\mu$,
  in Equation \eqref{eq:lagrange}. These two constraints in our KKT
  system are said to be \emph{qualified}\footnote{since both equality
    and inequality constraints are affine functions, the linearity
    constraint qualification holds and thus the existence and
  uniqueness are enforced},
  which guarantees that there
  exists a unique solution to the problem in Equation \eqref{eq:KKT}.
\end{remark}

\subsection{Numerical stability concerns and Chebyshev expansion
\label{sec:cheby}}

Now, note that our whole framework boils down to a matrix inversion.
The robustness of a linear system's solution is quantified by the
\emph{condition number}
of the invertible matrix. For a given invertible matrix $A$ in a
system $Ax=b$, the condition number, denoted
$\kappa(A)$, is defined as:

\[
  \kappa(A) = \|A\| \times \|A^{-1}\|,
\]
where $\|\cdot\|$ is a matrix norm. This value can be understoo as an
amplification factor for
errors. More specifically, it bounds the relative error in the solution
$x$ with respect to
the relative error in the input $b$. A large condition number signales an
\emph{ill-conditioned} system, where small numerical inaccuracies, such as
floating-point rounding errors inherent to computation, can be
magnified into large and erroneous deviations in the final solution.

In our optimisation, the KKT formulation of Equation~\eqref{eq:KKT}
requires solving a
system with a block matrix containing the term $2S$. The condition
number of this KKT
matrix can become prohibitively large, particularly as the number of
lenses $N$ increases
(e.g., for $N=5$ it is roughly $10^{13}$). The physical origin of
this ill-conditioning
is at the roots of the optical system: it arises when two or more
glasses in the lens system possess
very similar
dispersion characteristics.

The matrix $S = D_\mathrm{high} D_\mathrm{high}^\top$ encodes the
correlations between
the high-order chromatic contributions of each lens. If two glasses,
say lens $i$ and
lens $j$, have nearly identical dispersion curves, their
corresponding rows in the
$D_\mathrm{high}$ matrix will be almost linearly dependent.
Consequently, the $i$-th and
$j$-th rows and columns of the matrix $S$ will also be nearly
linearly dependent, driving
the matrix close to singularity and thus causing its condition number
to skyrocket. From
an optimisation perspective, the algorithm cannot reliably
distinguish between the
contributions of these two similar glasses, which can lead to unstable solutions
characterised by large positive and negative powers for $K_i$ and
$K_j$ that nearly
cancel each other out, a result that is mathematically plausible but
physically impractical.

To improve the computation of the matrix $S$, we could rather use an analytical
representation of the dispersion curves, $n_i(\lambda)$, for each lens
material, such as polynomial basis.

The Chebyshev polynomials \cite{Lanczos1952} are an excellent choice
for this task. As a
set of orthogonal polynomials, they provide a stable and efficient
representation of dispersion functions. Unlike
standard power series,
which can suffer from numerical instability (Runge's phenomenon)
especially at the edges of the fitting interval, Chebyshev expansions minimise
the maximum error and prevent wild oscillations. This stability is
crucial for ensuring that the resulting power function $\phi(\lambda)$
is well-behaved across the entire spectral band.

This robust representation is a critical first step, as it
significantly improves the
stability of the KKT solution compared to using a more naive
polynomial fit. However, it
is important to
distinguish between the accuracy of the physical model and the
inherent conditioning of
the optimisation problem itself. Even with a perfect representation
of the dispersion
curves, the matrix $S$ will remain fundamentally ill-conditioned if
the chosen set of
glasses contains materials with very similar chromatic properties. The Chebyshev
expansion provides a helpful stabilisation, but it cannot alter the
fact that the resulting
KKT system may be sensitive to small perturbations.

Hence, while employing a Chebyshev basis is a valuable method that
stabilizes the
problem to a certain degree, we believe it is also useful to
address the algebraic
structure of the solution directly. For this reason, we can also turn
to the null-space method
\cite{Gould2001}.
This approach is not mutually exclusive with using Chebyshev
polynomials; rather, it
takes the high-fidelity matrix $S$ they help produce and solves the constrained
optimisation problem in a way that is inherently robust to the very
correlations that
cause numerical instability in the KKT framework.

\subsection{Dealing with ill-conditioned optical system: the null-space method
\label{sec:null-space}}

Even with a Chebyschev expansion, the numerical condition number of
the inverted matrix
can still be really high, near $10^{10}$.
A more robust and as equally powerful approach to solving this
constrained optimisation
problem is the null-space method \cite{Gould2001}, a widespread
produce in computational
physics\footnote{known in optimisation as reduced 'reduced Hessian
  methods', structural
  mechanics as the 'force method', fluid mechanics as 'dual variable
  method' or electrical
engineering as 'loop analysis'.}. This is a complement of the latter
sections, as it
computes a stablier solution than a bare KKT.
However, note that it can \textbf{only} deal with \textbf{equality
constraints} such as
the one Equation \eqref{eq:sum-power}, on the contrary to simple KKT
as in section \ref{sec:kkt}.
Hence one cannot enforce an inequality constraint
as $K_\text{min} < K < K_\text{max}$ with this method, but it shows
really useful for
only equality constraint.

The core idea is to decompose the
problem into two distinct parts: one part that satisfies the required
mean power and
another that focuses exclusively on minimizing the chromatic variance
without affecting
the mean power.

First, we find any single solution, which we will call the
\emph{particular solution}
$K_p$, that satisfies the mean power constraint, $C_0^\top K_p =
\phi_0$. This can be
thought of as a baseline design, that has the correct overall power
but is not yet
optimised for chromatic performance.

Next, we characterise all possible modifications to this baseline
design that \emph{do
not change the mean power}. These modifications form the \emph{null
space} of the
constraint. Any vector $K_h$ in this null space satisfies $C_0^\top
K_h = 0$. For an
$N$-lens system with one constraint, there are $N-1$ ways
to adjust the lens
powers against each other while keeping the total power fixed. We can
group these
fundamental adjustments into the columns of a matrix $Z \in \R^{N
\times (N-1)}$.

Any optimal solution $K^*$ can now be expressed as the sum of our
baseline design and a
combination of these power-neutral adjustments:

\[
  K^* = K_p + Z y.
\]

Here, $y \in \R^{N-1}$ is a vector of coefficients that determines
the extent to which
each null-space adjustment is applied. Since the $Z y$ term, by
construction, does
not affect the mean power, the constraint is automatically satisfied.
Our task boils down
to finding the optimal coefficients $y$ that minimise the chromatic variance.

By substituting this expression for $K^*$ into the objective function
$K^\top S K$, the
constrained problem in terms of $K$ is transformed into a simpler,
\emph{unconstrained}
problem in terms of $y$:

\[
  \operatorname*{argmin}_{y} \mathcal{J}(y) = (K_p + Z y)^\top S (K_p + Z y).
\]

This is a standard quadratic minimisation problem similar to the one
explained before,
the solution is found by setting $\nabla_y \,\mathcal{J}$ the
gradient with respect to $y$ to zero, which yields a smaller and
well-conditioned system
of linear equations:

\[
  (Z^\top S Z) y = -Z^\top S K_p.
\]

As a recap:
\begin{itemize}
  \item $K_p$ represents a simple, initial lens configuration with the target
    power $\phi_0$.
  \item $Z$ is a matrix whose columns represent 'chromatic' modes;
    namely ways to alter
    individual lens powers without affecting the overall mean power.
  \item $y$ is the vector of weights for these modes.
  \item The matrix $Z^\top S Z$ is the 'reduced Hessian' from the
    KKT, it describes
    the chromatic variance within the subspace of power-neutral designs.
\end{itemize}

Solving this smaller system gives the optimal weights $y^*$. The
final, optimised lens
powers are then assembled through $K^* = K_p + Z y^*$. This method
effectively separates
the problem of achieving the target power from the task of color
correction, offering an
numerically robust tool for the optimal design. Indeed, the condition
number of the
inverted matrix finally bottoms down to $10^{5}$ instead of $10^{15}$
for a bare KKT in
section \ref{sec:kkt} or $10^{10}$ with Chebyschev expansion, and
even lower than the root-finding method in \ref{sec:n-achromat} at $10^8$.
\section{Numerical results}

As a reminder, we emphasis that both root-finding method (such as
superchromat/pentachromat) or optimal chromatic error method computes a set of geometric
factor $K_i$ that acts as a degree of freedom to improve the residual chromatic error on a bandwith.

\subsection{Pentachromat vs superachromat}

\begin{figure}[!t]
  \centering
  \includegraphics[width=\textwidth]{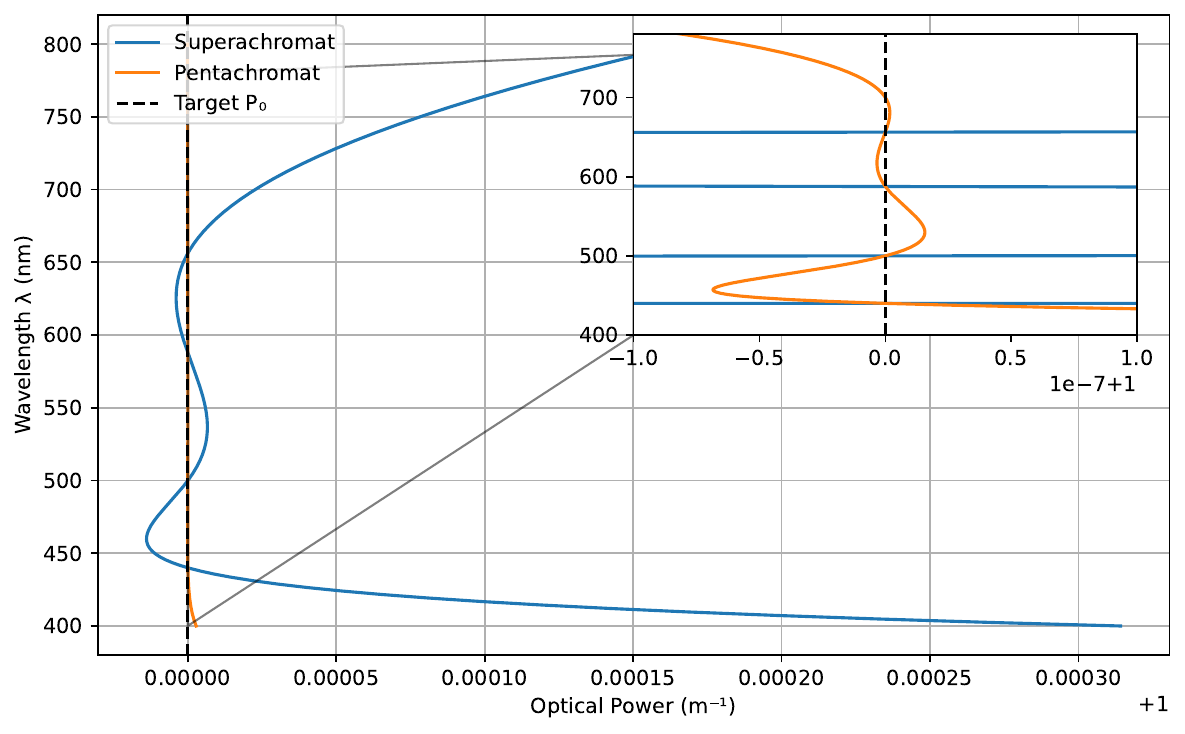}
  \caption{Comparison between pentachromat and superchromat,
    solutions to \ref{N_achromat} for $N=5$ and $N=4$ respectively.
    The deviation with respect to target value is reduced
  by 2 orders of magnitude when a pentachromat solution is used.}
  \label{fig:comparo-penta}
\end{figure}

Solving system \ref{N_achromat} with $N=5$ gives an analytical formula solving exactly,
using symbolic computation, the optical power for five different colors.
The corresponding formula is given in appendix.
Let us remark that a similar approach was proposed in \cite{Lu:22},
but in the case of a non cemented system. 
Both approaches have pros and cons, depending on the context (compactness, alignement ease, ...).
On Figure \ref{fig:comparo-penta}, we compare a superachromat lens, bringing

\begin{align}
  \lambda_i = (0.440,\quad 0.500,\quad  0.5876,\quad  0.6563) \quad \mu\text{m}
\end{align}

to the same focal point $\phi_0 = 1$m$^{-1}$. We rely on glasses from the Schott
catalogues, here N-BK7, N-F2HT, N-SF5, N-LAK22.
With a pentachromatic lens, bringing

\begin{align}
  \lambda_i = (0.440,\quad  0.500,\quad  0.5876,\quad  0.6563,\quad  0.700) \quad \mu\text{m}
\end{align}

to the same focal point.
The residual spectrum is reduced by several oders of magnitude when using a pentachromat
compared to the superachromat.

The extra lens N-BAF10 in the pentachromat allows it to beat consistently the
superachromat/$4$-achromat here on the Vis, at
the cost of an explosive behavior in the UV and IR bandwiths.

\subsection{Optimal achromatic $N$-lenses}

We now turn to the practical computation for our optimal design. The choice of numerical
optimisation algorithm (KKT in \ref{sec:kkt}, KKT with Chebyschev basis in
\ref{sec:cheby}, KKT with null-space in \ref{sec:null-space})
is ultimately a practical one, guided by the
problem's characteristics. For the $N=5$ lens system considered above, the
selected set of optical glasses is sufficiently distinct in its chromatic properties.
Consequently, the resulting KKT system is not 'too' severely ill-conditioned and can be
solved with the Chebyschev expansion, without relying on the null-space method.

Note that we choose the Chebyshev basis for the computation of the
KKT system in equation \eqref{eq:KKT}, but any orthogonal basis of
polynomials can do the trick (up to the numerical stability). For the
pedagogical sake, we will plot the polynomial obtained in Chebyshev
basis but also in the Legendre basis, another famous basis adapted
for slowly varying polynomials. The computed $K_i$ will be the same
for both basis, so we will not mention them.

For the case $N=4$, we invert the linear system to find the
superachromat coefficient, and invert the KKT system to solve the
optimal dispersion achromat described with a Chebyshev basis. The
computed coefficients are given in the following table:

\begin{table}[H]
  \centering
  \begin{tabular}{@{}lS[table-format=2.6]S[table-format=2.6]S[table-format=+1.6]S[table-format=+2.4]@{}}
    \toprule
    \textbf{Coefficients} & \textbf{Superachromat} &
    \textbf{Chebyshev} & {$\Delta K = K' - K$} & {$\frac{\Delta K}{K}
    \times 100$ (\%)} \\
    \midrule
    $K_1$ & 1.087414 & 1.208196 & +0.120782 & +11.1073 \\
    $K_2$ & -19.969685 & -19.057129 & +0.912556 & -4.5697 \\
    $K_3$ & 7.826501 & 7.401755 & -0.424746 & -5.4270 \\
    $K_4$ & 14.234056 & 13.588775 & -0.645281 & -4.5334 \\
    \bottomrule
  \end{tabular}
\end{table}

Exactly the same applies to $N=5$, yielding the following table

\begin{table}[ht!]
  \centering
  \begin{tabular}{@{}lS[table-format=2.6]S[table-format=2.6]S[table-format=+1.6]S[table-format=+2.4]@{}}
    \toprule
    \textbf{Coefficients} & \textbf{Pentachromat} &
    \textbf{Chebyshev} & {$\Delta K = K' - K$} & {$\frac{\Delta K}{K}
    \times 100$ (\%)} \\
    \midrule
    $K_1$ & 7.618286 & 7.647080 & +0.028794 & +0.3779 \\
    $K_2$ & -4.099110 & -4.021590 & +0.077520 & -1.8919 \\
    $K_3$ & 0.763709 & 0.727460 & -0.036249 & -4.7471 \\
    $K_4$ & 9.971334 & 9.955367 & -0.015967 & -0.1601 \\
    $K_5$ & -9.265021 & -9.309969 & -0.044948 & +0.4852 \\
    \bottomrule
  \end{tabular}
\end{table}

\begin{figure}[ht!]
  \centering
  \includegraphics[width=\textwidth]{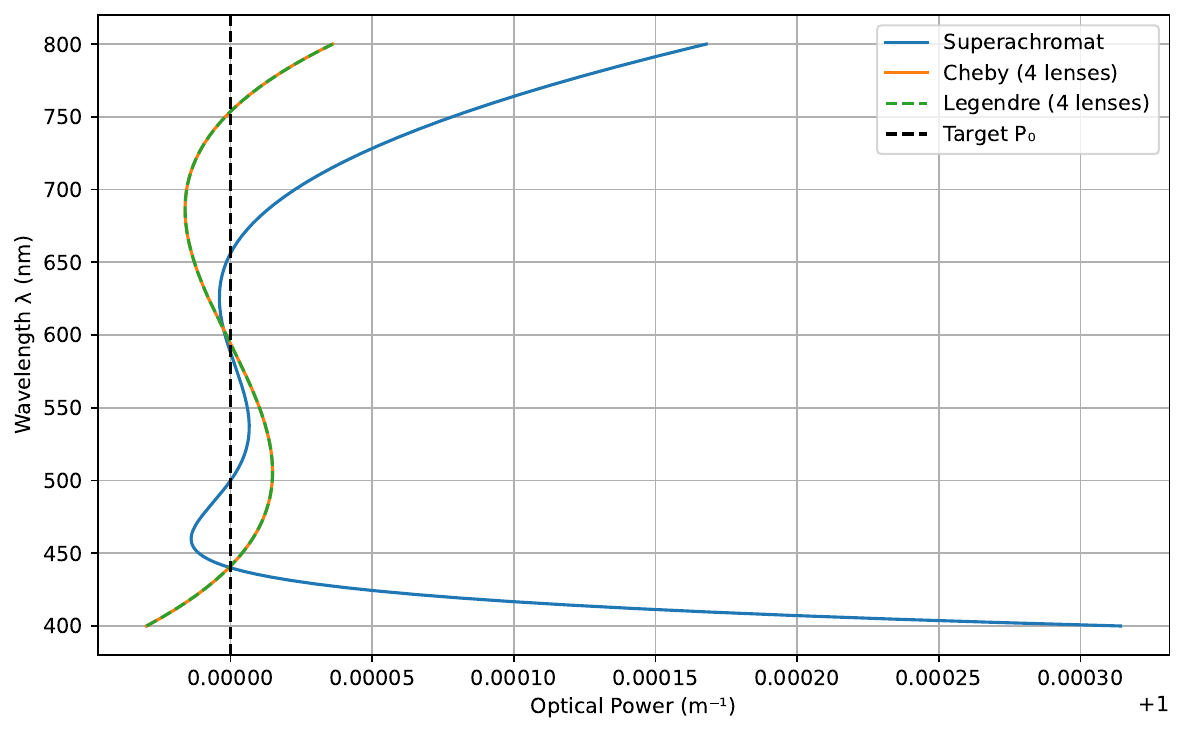}
  \caption{Optical power with respect to wavelength for the
    superachromat (four roots) and the optimal polynomials in either
    Chebyshev  and Legendre basis. The optimal polynomial clearly
    outperforms the superachromat, as its $\mu$m$^{-2}$ MSE is $\num{3.06e+1}$ $\mu$m$^{-2}$
  versus the superachromat $\num{2.12e+02}$ $\mu$m$^{-2}$.}
  \label{fig:super-vs-the-world}
\end{figure}

\begin{figure}[ht!]
  \centering
  \includegraphics[width=\textwidth]{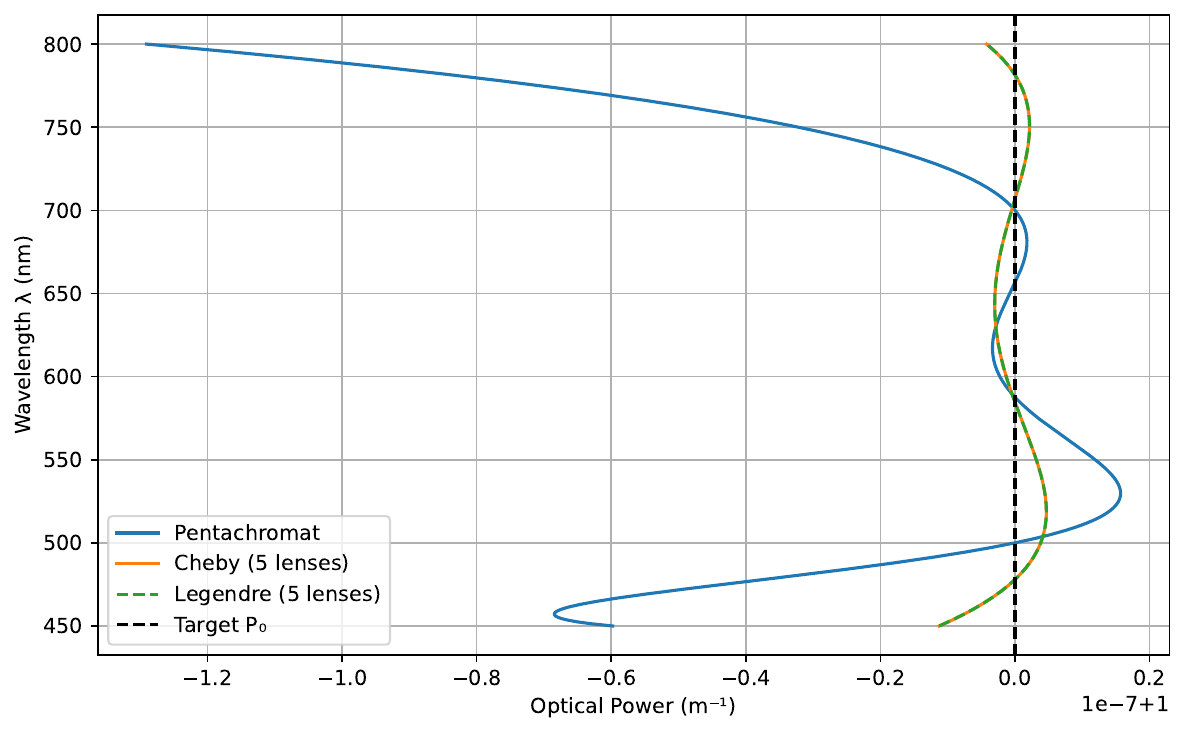}
  \caption{Optical power with respect to wavelength for the
    pentachromat (five roots) and the optimal polynomials in either
    Chebyshev  and Legendre basis. The optimal polynomial clearly
    beats the pentachromat as its $\mu$m$^{-2}$ MSE is $\num{2.60}$ $\mu$m$^{-2}$
  versus the superachromat $\num{7.73e1}$ $\mu$m$^{-2}$.}.
  \label{fig:penta-vs-the-world}
\end{figure}

One can observe that the coefficients are quite close for the
superachromat and the corresponding optimal $4$-lenses.
This is all the more true for the $5$-lenses case, which is kind of
obvious as the coefficients are already well-spaced.
We also remark that the five glasses solution reduces the optical power, 
and therefore leads to solutions that are easier to produce in term of manufacturability.
Finally, we can plot the optical power with respect to wavelength
for the case $N=4$ in Figure \ref{fig:super-vs-the-world} and for
$N=5$ in Figure \ref{fig:penta-vs-the-world}.

The optimal $K_i$ given by the KKT system in either Chebyshev or
Legendre basis consistently outperforms the superachromat over the
entire wavelength span. In practice, this comes at the cost of the
optical power behavior for $\lambda \to +\infty$, where the optimal
polynomial explodes, with a greater growth than the superachromat. As
it concerns wavelength far beyond the window defined for the optimisation (the visible
spectrum), this is not a concern at all. Moreoever, the optimisation was performed for a
bandwith $400 - 800\unit{\nano\metre}$, but it can be easily changed
to a more wide (resp. tighter) bandwith to match the design choice.

\begin{remark}
In terms of orders of magnitude, 
the solutions for geometric factorsobtained are in a compatible with realistic applications, especially microscopy.
\end{remark}

\subsection{Multiple focal achromatic design}

\begin{figure}[ht!]
  \centering
  \includegraphics[width=\textwidth]{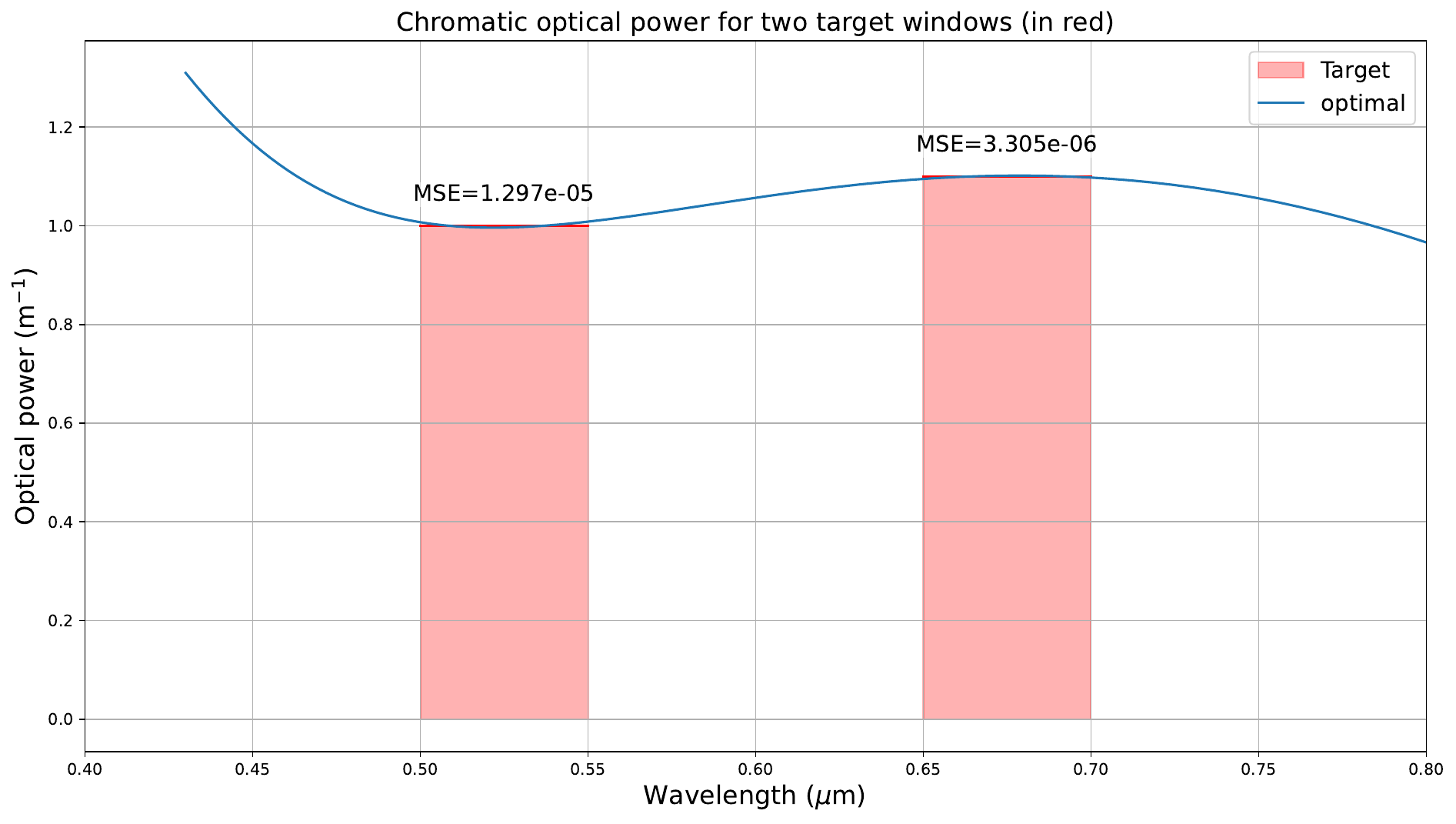}
  \caption{Dispersion function of the optimal polynomial matching a
    target power function on the red windows, with $\phi_0 = 1$
    m$^{-1}$ for 500 - 550 nm and $\phi_0 = 1.1$ m$^{-1}$ 650 - 700 nm.
  Root mean-squared error in m$^2$ is indicated for each windows.}
  \label{fig:target-windows}
\end{figure}

Our work can be easily generalised to a custom target power function
$t(\lambda)$ rather than the simple constraint in Equation \ref{eq:sum-power}.
The KKT system introduced in Equation \ref{eq:KKT} still holds, but
it should now encompass the difference between the target and the
chromatic error $K^\top S K$. This problem has a closed-form solution by
the normal equations. One can even mix mean-power $\phi_0$ and
wavelength-specific targets like the roots $\lambda_j$, by using a
Lagrange relaxation.
A peculiar case of interest is related to different optical power for different
bandwidths, like near-UV, visible spectrum (Vis) or near-IR. For instance, we could design a
target power of $\phi_0=1$ m$^{-1}$ and $\phi_0=1.1$ m$^{-1}$ for two different windows of the Vis.
Such an exemple of target windows application is given in Figure
\ref{fig:target-windows}.

\begin{remark}
Among the potential applications of such designs, 
let us note that this approach may be interesting in straylight and ghost reduction.
Indeed, changing the focal length for rays outise a given spectral window will have the effect to unfocus the ghosts.
\end{remark}

\subsection{Optimality on the glass choices}

The optimisation framework presented thus far computes the optimal curvature coefficients,
$K_i$, for a \emph{fixed} set of lens materials. However, a truly optimal design must
also consider the discrete choice of the glasses themselves. We extend our method to a
higher level of optimisation where the glass types are not inputs but variables. In a
practical scenario, a lens designer could provide a list of available materials, and the
system would identify the best combination. To demonstrate the power of this approach, we
performed an exhaustive search using the entire Schott optical glass
catalog\footnote{\url{https://www.schott.com/fr-fr/products/optical-glass-p1000267/downloads}},
thereby enabling the discovery of the globally optimal lens combination, both in the
continuous $K_i$ but also on the discrete set of glasses.

This combinatorial search is computationally intensive. For a system of $N=4$ lenses
selected from a catalog of over 122 materials, the number of possible combinations is on
the order of millions, as we have ${4\choose 122} = 8,783,390$ possibilities. To make this search
tractable, we also implemented a fully parallel beam search algorithm, which efficiently
explores the vast solution space to locate the most promising candidates. For each
potential combination of $N$ glasses, the optimal powers $K_i$ were computed using the numerically
robust null-space method, ensuring stability across all tests.

This exhaustive search\footnote{The full computation took 1 hour, while the beam search
took 30 seconds, both on a 2023 MacBook Pro M2 Max on power saving mode.} yielded two key
results, highlighting the
trade-off between
theoretical performance and practical manufacturability. The lens combination that
provides the absolute minimum residual chromatic error (MSE) was found to be:
\begin{itemize}
  \item \textbf{Best MSE Performance:} FK5HTi, N-LAK14, N-PK51, N-PK52A with an
    MSE of $9.11 \times 10^0$ nm$^{-2}$. The $K$ values are respectively:
    \[
      K =
      \begin{pmatrix}
        -10.2592 & 4.6166 & -54.3731 & 63.4275
      \end{pmatrix}
    \]
\end{itemize}
However, optimal mathematical solutions sometimes yield very large lens powers $K_i$
(strong positive and negative curvatures), which can be difficult and expensive to
manufacture, practical feasibility is therefore a critical design constraint. To address
this, we implemented a second selection criterion. From the top 100 combinations with the
lowest MSE, we identified the one that exhibited the minimum sum of squared lens powers
$\sum_i {K_i}^2$. This acts as a heuristic to favor solutions with gentler curvatures. This
approach yielded:

\begin{itemize}
  \item \textbf{Best Performance with Minimal Powers:} LF5, N-BAF10, N-FK51A,
    N-LAF33 with a slightly higher MSE of $4.07 \times 10^2$ nm$^{-2}$. The $K$ values are
    respectively:
    \[
      K =
      \begin{pmatrix}
        8.4582 & -16.4145 & 9.019 & 3.4256
      \end{pmatrix}
    \]
\end{itemize}

This two-step process provides a practical way to enforce manufacturability. As mooted
before, this constraint could also be formally integrated into the optimisation by
adding an inequality constraint on the magnitude of the $K_i$ values directly into the
Lagrangian, though our heuristic approach proves highly as effective and more simple in practice.

\section{Conclusion}

In this work, we have introduced a comprehensive and unified framework for the design of
cemented lens systems that improves considerably the correction of chromatic aberrations.

First, we derived a simple and unified framework to get a closed-form 
analytical solution for correcting $N$ colors.
Second, recognising the limitations of correcting only discrete spectral lines, we introduced a
more general variational framework. By minimising the integrated power variance over
a continuous spectral window, this approach guarantees a theoretically optimal solution
for any given set of materials. We demonstrated how this Lagrangian-based optimisation,
when implemented with numerically robust techniques such as the null-space method,
provides a stable and reliable tool to yield the optimal design, even for ill-conditioned
systems where traditional methods may fail.

Despite covering paraxial optics, this work did not cover explicitly field curvature and Petzval sum, 
as it is often done when looking for curvatures of optical systems. 
This can be easily achieved, by using a symetrization strategy, 
such as the one done in Double Gauss lenses types (Clark, Planar...).
This solution elegantly circumvent the problem, as not only the Petzval curvature is zero on a selected number of wavelengths, 
it is true on a continuum (and then, in the spirit of reducing the RCA instead of LCA/TCA).
Also, aberration theory is not covered in this work: this may be done by using a classical 
approach of lens bendings (\cite{Schwarzschild1905}).

The most interesting aspect of this paper lies in its generalisability, which
transforms lens design from a series of local optimisations into a global, automated
process. The framework is not constrained to a fixed number of lenses or a predefined set
of glasses. By coupling our variational problem with a parallelised combinatorial search,
we created a design procedure capable of exploring millions of material combinations from
entire manufacturer catalogs to identify the globally optimal design. This process
balances the trade-off between peak theoretical performance (minimum MSE) and practical
manufacturability (minimum lens powers).

Ultimately, this methodology opens uncharted territories in optical design. It is not
limited to a single continuous band of interest; it can be leveraged to simultaneously
optimise performance across multiple, disjoint spectral windows, even assigning different
target focal lengths to each.

\newpage

\appendix


\section{Pentachromat: analytical solution}

Given $n_i(\lambda_p)$ the refractive index of lens $i$ at wavelength 200nm $<\lambda_p<$ $2\mu$m:

\[
  \begin{cases}
    u_{11} = n_1(\lambda_1)-1,  \\
    u_{12} = n_1(\lambda_2)-1,  \\
    u_{13} = n_1(\lambda_3)-1,  \\
    u_{14} = n_1(\lambda_4)-1,  \\
    u_{15} = n_1(\lambda_5)-1,  \\
    l_{21} = \frac{n_2(\lambda_1)-1}{u_{11}},  \\
    l_{31} = \frac{n_3(\lambda_1)-1}{u_{11}},  \\
    l_{41} = \frac{n_4(\lambda_1)-1}{u_{11}},  \\
    l_{51} = \frac{n_5(\lambda_1)-1}{u_{11}},  \\
    u_{22} = (n_2(\lambda_2)-1) - l_{21} u_{12}, \\
    u_{23} = (n_3(\lambda_2)-1) - l_{21} u_{13},  \\
    u_{24} = (n_4(\lambda_2)-1) - l_{21} u_{14},  \\
    u_{25} = (n_5(\lambda_2)-1) - l_{21} u_{15},  \\
    l_{32} = \frac{1}{u_{22}} \left( (n_2(\lambda_3)-1) - l_{31} u_{12} \right),  \\
    l_{42} = \frac{1}{u_{22}} \left( (n_2(\lambda_4)-1) - l_{41} u_{12} \right),  \\
    l_{52} = \frac{1}{u_{22}} \left( (n_2(\lambda_5)-1) - l_{51} u_{12} \right),  \\
    u_{33} = (n_3(\lambda_3)-1) - (l_{31} u_{13} + l_{32} u_{23}),  \\
    u_{34} = (n_4(\lambda_3)-1) - (l_{31} u_{14} + l_{32} u_{24}),  \\
    u_{35} = (n_5(\lambda_3)-1) - (l_{31} u_{15} + l_{32} u_{25}),  \\
    l_{43} = \frac{1}{u_{33}} \left( (n_3(\lambda_4)-1) - (l_{41} u_{13} + l_{42} u_{23})
    \right),  \\
    l_{53} = \frac{1}{u_{33}} \left( (n_3(\lambda_5)-1) - (l_{51} u_{13} + l_{52} u_{23})
    \right),  \\
    u_{44} = (n_4(\lambda_4)-1) - (l_{41} u_{14} + l_{42} u_{24} + l_{43} u_{34}),  \\
    u_{45} = (n_5(\lambda_4)-1) - (l_{41} u_{15} + l_{42} u_{25} + l_{43} u_{35}),  \\
    l_{54} = \frac{1}{u_{44}} \left( (n_4(\lambda_5)-1) - (l_{51} u_{14} + l_{52} u_{24}
    + l_{53} u_{34}) \right) \\
    u_{55} = (n_5(\lambda_5)-1) - (l_{51} u_{15} + l_{52} u_{25} + l_{53} u_{35} + l_{54} u_{45})\\
    y_1 = \phi_0 \\
    y_2 = \phi_0 - l_{21} y_1 \\
    y_3 = \phi_0 - l_{31} y_1 - l_{32} y_2 \\
    y_4 = \phi_0 - l_{41} y_1 - l_{42} y_2 - l_{43} y_3 \\
    y_5 = \phi_0 - l_{51} y_1 - l_{52} y_2 - l_{53} y_3 - l_{54} y_4 \\
    K_5 = \dfrac{y_5}{u_{55}} \\
    K_4 = \dfrac{y_4 - u_{45} K_5}{u_{44}} \\
    K_3 = \dfrac{y_3 - u_{34} K_4 - u_{35} K_5}{u_{33}} \\
    K_2 = \dfrac{y_2 - u_{23} K_3 - u_{24} K_4 - u_{25} K_5}{u_{22}} \\
    K_1 = \dfrac{y_1 - u_{12} K_2 - u_{13} K_3 - u_{14} K_4 - u_{15} K_5}{u_{11}} \\
  \end{cases}
\]

\bibliographystyle{plain}
\bibliography{biblio}

\end{document}